# Rediscovering Practice and Inquiry in Academic Education: Experiences in a European University Environment

Sebastiano Cantalupo

Department of Physics, University of Milan Bicocca, Milan, Italy
Department of Physics, ETH Zurich, Zurich, Switzerland
sebastiano.cantalupo@unimib.it

## Abstract

I describe the design and implementation of a series of university MSc courses in Switzerland and in Italy on the topic of "Cosmic Structure Formation" whose goal has been to provide to the students a *formative* experience using interwoven research *practice* and fundamental scientific content. The course educational framework, which is based on the ISEE Inquiry Framework, emphasizes science, as much in teaching as in research, as a set of *practices*, re-discovering and actualizing in modern terms the original pivotal role which these practices had in education in ancient times. In particular, the courses focus on *formative, intuitive, student-centered* and *dialogic* learning in opposition to the *informative, mnemonic, teacher-centered* and *monologic* teaching of frontal lecture-based instruction, which is still the dominant teaching framework in university education, at least in Europe. I describe how course activities are designed in such a way as to mirror authentic research, including all aspects which are usually not *practiced* in lecture-based courses and "standard" laboratories (e.g., generating and refining questions; making and testing assumptions; developing one's own research path; and sharing, explaining and justifying ideas and results with peers). Finally, I discuss the major outcomes of the courses and the main challenges which were faced in order to provide to the students a truly *transformative* experience which could allow them to improve both as learners and future scientific researchers, as well as members of a larger community.

Keywords: cosmic structure formation, course design, inquiry, STEM practices

## 1. Introduction

" 'Come and listen to me read my commentaries. I will explain Chrysippus to you like no one else can, and I'll provide a complete analysis of his entire text…' So it's for this, is it, that the young are to leave their homelands and their parents: to come and listen to you explain trifling little words?" (Epictetus, *Discourses*, Book 3, Discourse 21)

This is the voice of one of the most famous teachers of his age — the stoic Epictetus — as recorded by one of his students in his classroom in Nicopolis, in eastern Greece, at the beginning of the II century CE. Part of a long tradition dating back to Plato's Academy and the teaching of Socrates in the V to IV century BCE, Epictetus' view on the importance of "practice" in teaching and learning are shared among the majority of schools of what we call "classical antiquity" in Europe, as demonstrated by the seminal works of philosopher and historian of philosophy Pierre Hadot (1995, 2001, 2004). Including a curriculum based on physics, ethics and logic, these schools taught *philosophy* as intended in "its original aspect: not as a theoretical construct, but as a method for training people to live and to look at the world in a new way" (Hadot, 1995, p. 107). Physics, intended in its original meaning of "nature" or study of "nature" (from the Greek *physis*), occupied a very prominent role in the teaching of almost all ancient schools, and in particular for the Stoics for which "the parts of philosophy [physics, ethics, logic] are equal and mutually imply one another: to practice one of them is necessarily to practice all of them" (Hadot, 2001, p. 79). These three themes were interwoven and represented our relationship and our place within the universe or "nature" as a whole (physics), within our human or particular community (ethics) and within ourselves in terms of our inner discourse (logic). In the words of the Stoic Emperor Marcus Aurelius (II century CE) written in his notes, or *exercises,* for himself:

> "Continuously, and if possible, in every occasion apply to your thoughts *physics, ethics and logic*" (Marcus Aurelius, *Meditations,* Book VIII, Meditation 13); "This you must always bear in mind, what is the nature of the Universe, and what is my nature, and how this is related to that, and what kind of a part it is of what kind of a whole" (*ibid.,* Book II, Meditation 9).

Training and education in these disciplines, in antiquity, "was still, fundamentally, a dialogue. The goal was not to *inform,* or to transplant specific theoretical contents into the students' minds, rather, it was to *form* them" (Hadot, 1995, p. 87). The same applies also to the few cases in which a written form of dialogue has been used: the most famous examples are Plato's Dialogues, whose structure, different than what we would modernly call a systematic treatise, is suggestive of their *formative* rather than *informative* goal (Goldschmidt, 1963). These written dialogues were not intended in any case to substitute oral teaching in form of open discussion and debates within the "school" (Hadot, 1995). *Dialogic,* oral teaching responded perfectly to the nature of ancient schools, which were mainly living communities of learners and peers who shared the common interests of the search and love for wisdom (*philosophia*). The dialogue was the *practice* to train, through questions and inquiry, the inner discourse which the learners use for their own learning process: "Thought and dialogue are the same thing, except that it is the silent inner dialogue with ourselves which we have called 'thought'" (Plato, *Sophist*, Section 263e, 4).

However, like every method, dialogue and discussion also have their limitations. In addition to the limits of language, which cannot express the totality of reality and is limited by words and "definitions", continuous *practice* is in any case required: "Those who have begun to learn link words together but do not yet know their meaning; for the words must be *integral parts of our nature*. But this takes time" (Aristotle, *Nicomachean Ethics*, Book VII, 1147a21-22). In ultimate analysis, as shown by P. Hadot, true learning for the schools of classical antiquity in the Greek, Hellenistic and Roman periods in Europe corresponded to a deep transformation of the self, which required continuous practice and effort and the active participation of the learner.

This attitude towards the learning process shares many resemblances with many schools that developed in India and in the Far East in the same period (VI century BCE to the II century CE) and in particular with Daoism in China. In the collection of short stories and sayings which are attributed to Master Zhuang or Zhuangzi (莊子) of the III century BCE, there are many examples in this regard.

It is instructive to report here a couple of these anecdotes.

> "Duke Huan was in his hall reading a book. The wheel-wright Pian, who was in the yard below chiseling a wheel stepped up into the hall, and said to Duke Huan, 'This book Your Grace is reading—may I venture to ask whose words are in it?' 'The words of the sages,' said the duke. 'Are the sages still alive?' 'Dead long ago,' said the duke. 'In that case, what you are reading there is nothing but the chaff and dregs of the men of old! […] I look at it from the point of view of my own work. When I chisel a wheel, if the blows of the mallet are too gentle, the chisel will slide and won't take hold. But if they're too hard, it will bite and won't budge. Not too gentle, not too hard— you can get it in your hand and feel it in your mind. You can't put it into words, and yet there's a knack to it somehow. I can't teach it to my son, and he can't learn it from me. So I've gone along for seventy years, and at my age I'm still chiseling wheels. When the men of old died, they took with them the things that couldn't be handed down.' " (Zhuangzi, *Complete Works*, Book 13, Fragment 7)

"Confucius said to Laozi [the founder of Daoism], 'I have been studying the Six Classics for what I would call a long time, and I know their contents through and through. But I have been around to seventy-two different rulers with them, expounding the ways of the former kings and making clear the path trod by the dukes of Zhou and Shao, and yet not a single ruler has found anything to excite his interest. How difficult it is to persuade others, how difficult to make clear the Way!' Laozi said, 'It's lucky you didn't meet with a ruler who would try to govern the world as you say. The Six Classics are the old worn-out paths of the former kings — they are not the thing that walked the path. What you are expounding are simply these paths. Paths are made by shoes that walk them; they are by no means the shoes themselves!' " (Zhuangzi, *Complete Works*, Book 14, Fragment 7)

As for the ancient Greeks, also for the Daoist the words are useless unless they are lived and practiced. Their teaching did not require systematic treatises expounded by a master in front of an audience. For them, the path must be walked to be learned. Words could still be used as a *facilitating* device for the beginner, but eventually, when they became *integral parts of our nature* (as the ancient Greeks would say) they could be forgotten:

> "The fish trap exists because of the fish; once you've gotten the fish, you can forget the trap. The rabbit snare exists because of the rabbit; once you've gotten the rabbit, you can forget the snare. Words exist because of meaning; once you've gotten the meaning, you can forget the words. Where can I find a man who has forgotten words so I can have a word with him?" (Zhuangzi, *Complete Works*, Book 26, Fragment 14)

It is outside of the scope of this Introduction to discuss similarity in the learning approach of other ancient cultures around the world or to provide a complete historical overview. However, it is interesting to notice that while in the Far East and in other cultures, *practice* remained a focus of teaching, a radical shift happened in Europe after the fall of the Roman Empire in correspondence to deep changes in the religious and social context. The classical schools were closed and knowledge was confined to monastic environments. This paradigm shift in teaching methods propagated through the Dark and Middle Ages — when the first universities were born — until the present epoch (e.g., Hadot, 2004). In particular, the central role was taken by authoritative texts, paradoxically from the same ancient authors discussed above: the *dialogue* became a *monologue* of the teacher, whose main role was to interpret the text. Authoritative text existed also in the "classical" period, but as we have heard from the voice of Epictetus at the beginning of this introduction, their exegesis (or *lectio* from the Latin "to read") was not the central part of his teaching. The *lectio*, from which the modern English word "lecture" and "lesson" is derived, became instead the central and often only part of the teaching curriculum of the first universities in Europe until today. The Scientific Revolution of the XVI and XVII centuries CE and the rise of the scientific methods disconnected research activities from the simple exegesis of authoritative texts. This was not followed, however, by a similar revolution in university teaching. Moreover, the historical and religious context of that time did not allow the study of nature to regain its role of a *formative* or *transformative* learning experience as it was in antiquity: "He who does not know that the Universe exists, does not know where he is." (Marcus Aurelius, *Meditations*, Book VIII, Meditation 52).

In this article, I describe my personal experience in the design and implementation of a series of university MSc courses for graduate students in Switzerland and in Italy during the last five years on the topic of "Cosmic Structure Formation." The goal of these courses was to provide to the students with a *formative* experience using interwoven research *practice* and content related to the study of the Universe. The educational framework for the design of these courses took advantage of several decades of works and studies by educators which emphasizes science, as much in teaching as in research, as a set of *practices* (see e.g., Metevier et al., 2022a, in this collection). These studies re-discovered and actualized in modern terms the original pivotal role of *practice* in learning in ancient times, as we have seen above, although their focus was mostly on elementary, high school, and undergraduate students, rather on the graduate level. Often referred to as "inquiry" in the literature, this framework focuses on *formative, intuitive, student-centered* and *dialogic* learning in opposition to the *informative, mnemonic, teacher-centered* and *monologic* teaching of frontal lecture-based instruction, which is still the dominant teaching framework in university education, at least in Europe. In the inquiry framework, the frontal teacher becomes a *facilitator* whose role is not to transfer ready-made knowledge but rather to guide and help the learners travel their own path through their own learning process. For inquiry, as much as for the schools of classical ages in Europe and in the Far East, words are useless unless they are *practiced*: the path must be walked to be learned.

In the context of the Inquiry Framework (Metevier et al., 2022b) developed by the Institute for Scientists & Engineer Educators (ISEE), on which my course designs have been based, *practice* is intertwined with foundational concepts, although for practical reasons they are separated in the course design and assessment of learners. *Practice* is designed in such a way as to mirror authentic research, including all aspects which are usually not *practiced* in a traditional laboratory class (e.g., generating and refining questions; making and testing assumptions; developing one's own research path; and sharing, explaining and justifying ideas and results with peers). Finally, the students develop ownership of the learning process, including generating their own evidence to support an explanation of their understanding.

The article is structured as follows. In Section 2, I describe the context in which these courses took place in terms of the students' prior learning experiences. In Section 3, I provide an overview of the course design in light of the three main themes of the ISEE framework (Inquiry, Equity and Inclusion, Assessment). In Section 4, I discuss the outcomes, successes and limitations in the context of the courses and their environment. A summary is provided in Section 5.

## 2. Context

The course on the topic of "Cosmic Structure Formation", with a typical duration of about 40 hours,

took place during several academic semesters at the Federal Institute of Technology Zurich, Switzerland (ETH) from 2017 to 2019 and at the University of Milan Bicocca, Italy (UniMiB) in 2020 and 2021 under different names. The content and practice side of the course has been continuously updated in order to reflect lessons learned during the previous years, as described in detail in Sections 3 and 4. The course has been part of the MSc in physics at ETH from 2017 to 2019 (offered for 4 semesters) but open to all students including Bachelor's and Doctoral students, including students with backgrounds different than physics. On average, between 8–13 students attended the courses (which is a typical number for "optional courses" at the MSc in physics at ETH and UniMiB). The majority of them had a background in physics, followed by students with a background in mathematics or computer science and including also in some instances students from different fields such as material sciences. In addition to a large variety of career levels, the students had also very different personal backgrounds: the participants typically included students from various parts of Europe. The largest fraction of the remaining students originated from China. The majority of the students in the courses had previously received a BSc at ETH or elsewhere in Europe with just a few cases of exchange students (from Australia and Japan). The participants in the courses which took place as part of the MSc in Astrophysics at UniMiB in 2020 (as a part of the "Laboratory of Astrophysics") and 2021 (about 20 and 13 students, respectively) had instead a much more homogenous background: they were all Italians with a background in physics, with the exception of two exchange students from elsewhere in Europe in 2021. In all cases, the course language was English. No students from the US (or UK) or with any prior experiences in the US attended the courses.

In the following, I give a brief and by no means exhaustive overview of the prior learning experiences both in terms of content and practice of a typical student attending the courses on "Cosmic Structure Formation". Both in Switzerland and in Italy, the students follow a 3-year BSc in physics, which is mostly based on frontal lectures that take place in large lecture rooms. They attend basic courses in mathematics and physics at ETH and UniMiB with a typical audience of 300 or more students, which reflects the large number of enrolled students in the BSc program (there are no enrollment constraints in Switzerland and Italy for physics BSc). The primary focus of these courses is to cover a large amount of content in a short time using a lecture-based approach, as discussed in the Introduction. During these courses, interaction between lecturer and students is typically limited to questions at the end of the lecture in the plenary session (when time allows), or individually with the teacher during office hours. The exposition of the content during the lecture follows a somewhat rigid structure based on a particular book or on lecture notes of the lecturer. The lecture notes are usually made available to the students, in most cases even before the lectures themselves. Together with the lecture part, the students (at ETH) might attend sessions where experiments related to the lecture content are shown by the lecturer to the audience.

For laboratories and exercise classes, the students are divided into smaller groups. In both cases, the students are typically presented with a set of highly structured tasks to solve, which may require applying a particular formula heard during the lecture. These tasks often have only one possible path and one possible outcome which constitute the "right" or "wrong" result. Assumptions are usually listed and given in the exercise itself. A few weeks before the exam, the students typically memorize as much as possible of the lecture content (provided by the lecture notes) which is then repeated back to the lecturer. As a result, a large fraction of the material presented in the courses is not ultimately retained. On the practice side, the students are mostly involved in solving ready-made tasks which could require significant mathematical skills but are far from an authentic research experience.

Drop rates after the first year are significant, either forced by *selection* procedures (e.g., the requirement to obtain a certain number of credits in a given time) or by voluntary abandons. This is seen not necessarily as a bad thing by a somewhat large fraction of the university and lecturer board, which identifies as a primary goal of large universities the *selection* of the students (according to some metrics) rather than the *formation* of the students. Such "selection" is performed on a very limited set of skills, mostly based on mathematical or abstract knowledge, and on short-term memory retention. The resulting effects on the diversity of the student population reaching the MSc are clearly noticeable both in terms of student gender, background and skills. More subtly, such a path affects the perception of students about scientific research (which they often erroneously associate with their BSc courses) and their self-perception as potential scientists.

## 3. Course design

Within the general context discussed in the previous Section, the courses on "Cosmic Structure Formation" have been designed with the goal of being a *formative* and eventually *transformative* learning experience for the students as well as the teacher, or facilitator. Given the previous learning experiences of the students in their previous academic courses (mentioned in Section 2), such a formative path should necessarily be a process of *rediscovery* of both fundamental physics content and practice. For an authentic learning experience, the *rediscovery* or "new discovery" must be a personal experience which is practiced and *lived* by the students themselves as discussed in detail in Section 1. The ISEE Inquiry Framework offers a set of practices and teaching strategies which are ideally suited to reach these goals, and the course design has been implemented taking advantage of this framework in which foundational scientific content and practice are interwoven.

In order to reach the goal stated above, the course is designed as a set of *dialogic* experiences rather than lecturer monologues. Here dialogic is used in the sense of a living dialogue, in contrast to dialectic. Dialogue is designed on three different levels which are interwoven but described here separately and in a sequential form for simplicity.

First, there is the dialogue between the facilitator and each individual student (or small group of students) in the form of facilitator's questions which have both the aim to guide the students through their learning experience and to train them in developing dialogues on the other levels. In this dialogue the facilitator is not the owner of knowledge. The questions are not aimed at obtaining an answer which is "right" or "wrong" (as in traditional final exams), they are not aimed at obtaining an answer at all, unless this may be useful for assessment. Similarly, the facilitator does not directly answer students' questions; rather, through questions, the facilitator shows possible ways in which the students could address their questions on their own.

Second, there are the dialogues between student peers, in small groups. Through these dialogues, students can learn by explaining. They may have an initial grasp of an idea or concept, an intuition. By putting them in words for others, they may consolidate their understanding, realize a possible error, or receive new ideas.

Third, there is the inner dialogue of each student, which is the level at which the learning process is consolidated and, in a sense, the end goal of the *dialogic* experience. The inner dialogue may be put in visible form for the facilitator and peers through the notes that students write in their notebooks, through sketches or diagrams, restarting the dialogic cycle on the other levels.

Finally, the learning experience is reported to a larger audience, i.e., to the whole classroom (or, in research, to the scientific community): this is the *sharing* phase, which can also be not dialogic but in the form of a presentation or written report. After this stage, another cycle can start.

In all these activities, "words" are the necessary means at all levels to make the learning process happen. They are connected and become meaningful through a logical process. A dialogic experience is therefore necessarily a *logical* exercise. Moreover, because a dialogue involves other people, including their opinions and values, a dialogic experience has necessarily also an *ethical* aspect and value.

In the following, I describe more in detail the central aspects of the course design in terms of foundational scientific content and practice. For the sake of exposition (with all its limitations), in this article, these interwoven sides of the learning process are described separately.

## 3.1 Foundational scientific content

A *transformative* experience is an experience that allows us to see the "world in a new way" (in the words of P. Hadot). Recent scientific and technological advancements give us the means to observe and study the extremely small of the atomic and subatomic world and the extremely large of the universe. Both these worlds are far away from our personal everyday experience and have thus the power to enlarge our view and the potential to transform it completely. These three apparently distinct worlds (atomic physics, our everyday world, the distant universe) are deeply interconnected.

One of the foundational content goals of the course was to let the students (re-)discover these connections. In particular, atomic-scale processes are the sources of electromagnetic radiation, which allows us to observe and study the whole universe. In addition to its messenger role, radiation is also an active agent, shaping the properties of the majority of structures in the universe: by losing energy through radiation, gas can cool down, and form galaxies, stars and planets. At the same time, the atomic and distant universes taken together are ideal benchmarks to let the students (re-)discover our (necessarily) limited view of the world, challenging their prior knowledge based on other courses or everyday experiences. For instance, the same object in the universe, as seen at different wavelengths of the electromagnetic spectrum, may appear completely different, revealing, for instance, the presence (and prevalence) of matter outside of galaxies. The properties of matter as seen through radiation can also reveal the necessity of other forms of ("dark") matter whose nature is still completely unknown. All these experiences can be used to elicit our *ignorance* rather than *knowledge* of the world around us, the authentic motor of pure inquiry and research.

In order to reach these goals within the course, a particular content framework has been chosen: the largest structures in the universe (Galaxy Clusters and the IntraCluster Medium; the Intergalactic Medium [IGM] and the Cosmic Web; the Early Universe and first structures; Galaxy Formation). Within this framework, through a set of facilitated activities, the students have the possibility to explore different scales and epochs in the history of the universe, as well as different regions of the electromagnetic spectrum and thus different atomic processes.

To facilitate final assessment the overarching content learning outcomes are summarized as follows in the course design: i) by using radiation processes on atomic scales, the students will learn how to derive the physical properties of the largest baryonic structures in the universe (Clusters of Galaxies, IGM, first structures in the Early Universe); ii) by using radiation processes on atomic scales compared with physical processes on cosmic scales, the students will learn that radiation is an active agent in cosmic structure formation and evolution.

A schematic representation of the course "Content Framework" is presented to the students at the beginning of each class providing a sort of *map* for their journey of discoveries. The division in three topics represents a good balance within the time available in a semester for a 6-credit course between deepening and enlarging the view. Exploring multiple topics allows the students to remain engaged and motivated. On the other hand, too many topics in a short time would not allow the necessary time for the students' discoveries and their "digestion".

The particular order of topics has been chosen to give to the students the sense of a journey in space and time. Indeed, because of the finite speed of light, astronomers are able to directly study the universe's past by looking at objects at larger distances (which we see as they were in the past, when they emitted the light we see now). Different radiation windows are investigated, from higher-energy radiation to lower energies, while moving to larger distances from us.

The activities designed for each topic have the goal of letting the students make "unexpected" major discoveries on their own without requiring complicated calculations or data analysis. These include the discovery of "dark matter", the "Cosmic Web" and the "Reionization of the Universe". At the same time, the activities allow the students to *re-discover* fundamental concepts and ideas which the students have encountered (and in most cases not assimilated) in previous courses, including the nature of light, several aspects of quantum physics, and fundamental concepts of thermodynamics and probability distribution functions.

## 3.2 Scientific practices

A set of practices has been chosen as core elements for the course design. This particular choice, described in detail below, is based on the expected needs of the students, considering both their prior experiences in courses (see Section 2) and the ISEE Themes.

### 3.2.1 Generating and refining questions

What is the starting point of scientific inquiry? How can we help the students to start their journey through their own learning process? There are of course multiple answers to these questions. The important fact, however, is that questions can start a process, which could be a research activity, a dialogue between different people or an inner dialogue which eventually can lead to learning. Inquiry learning takes advantage of this starting point, which makes learning more similar to an authentic research experience.

In order to stimulate the students to practice question-raising as a driver of their learning process through the course, a specific activity ("generating and refining scientific questions") is designed and offered as the introductory experience on the first day of the course. In addition to practicing question-raising, this activity has several goals which touch on multiple elements of the ISEE Framework, and the activity uses, at least in its initial part, some elements of the "Light and Shadows" activities used in ISEE's Professional Development Program (described in Hunter et al., 2010) and similar activities designed for the WEST workshops at University of California, Santa Cruz, in 2013 and 2014 (see e.g., Santiago et al., 2022, in this collection, for a description of WEST workshops).

After a brief plenary introduction session, material related to the content of the whole course is shown, including images of Galaxy Clusters at different wavelengths, spectra of distant quasars, and images of quasar fields at particular wavelengths. The material is just described as it is, no physical information or content is given. The goal of the introduction is only to stimulate interest and questions, which can be of any kind at this stage. The students then write questions (in a completely anonymous way) which are collected and shared with the whole class. The initial set of questions *is* the material of the facilitated activity which is conducted in small groups (formed randomly) including at maximum four students. The goal of the activity is to *refine* the question. Indeed, while all questions are fine, not all questions are equally investigable through scientific inquiry from a practical point of view.

After choosing one "unrefined" question, each group of students, through facilitation or with the help of a prompt, identifies the key elements which make a question more investigable from a scientific point of view. These include: i) the subject of the question is a clearly specified physical or observable variable which can be *measured* (or a relation between them), ii) the question is or can be turned into a testable statement, iii) if more subjects are present, the questions can be split into smaller units. The students then apply in practice these ideas and share their results in a plenary session.

Several facilitation strategies can be used to help the students achieve the goal of the activity. For instance, the facilitator could ask the students where they would start from *in practice* to address the question, and if they realize that they are in doubt, then likely the question can be further refined. A properly refined question is indeed a question that can easily start and drive a scientific investigation. Through the activity, the students also encounter for the first time several other core practices such as "defining a physical variable" (and distinguish it from an "observable variable"), "splitting complex problems in smaller units" and "sharing", which are described below.

Because the material for the activity is produced by the students themselves, this practice helps the students take ownership of the activity. The anonymous nature of the questions and the freedom to choose any question for the refining process are key elements for the Equity and Inclusion aspects of the activity. Moreover, the students take advantage of this opportunity to get to know each other without biases since groups are formed randomly (this is always the case for all the activities in the course). By generating and refining questions on material representing the Content Framework of the whole course, the students also take ownership of the whole learning process, which will also be guided by questions in the subsequent course activities.

### 3.2.2 Finding relevant physical variables

Considering the whole universe, or even our immediate surroundings, at once is an impossible task for the human brain. To relate to the external world, we typically operate a process of differentiation, which consists of defining individual objects and associating them with some particular attributes or characteristics. We can then process this "representation", relating different attributes and possibly finding unexpected relations which then allow us to learn new things. Representations are however not unique and they are always just a limited view of what they represent. For instance, a map is not, in a strict sense, the territory which it represents, and there are many ways in which we can represent the same patch of land for different purposes. As the ability to interpret a map and its symbols is fundamental to not get lost on a long journey in unknown territories, also the way in which we represent the world around us may facilitate a scientific investigation. An effective representation is fundamental for the three dialogic levels discussed above, allowing the facilitator and students to effectively interact which each other and with themselves.

In the context of the radiation phenomena which are studied and used in the course, the students are first trained in identifying "observable" variables in all the datasets and in "observing" things as they are without jumping right away into (physical) interpretation. For instance, it is expected that when presented with an image of the X-ray emission from the Intra Cluster Medium (ICM) of a Galaxy Cluster, the student will identify "intensity of radiation" as an observable variable. Or, when presented with an X-ray spectrum of the ICM, they would identify "specific intensity of radiation" and "radiation wavelength or frequency" as observable variables. More familiar sources are also used for analogies, e.g., the Sun. The observable variables in this case could be "intensity" and "color" of the radiation coming from the Sun. An observable variable is thus defined as a directly measurable quantity. The activities are then facilitated in such a way that the students move from the observations to the identification of the underlying, relevant physical variables. This often requires developing a "physical model" of the system. The "physical model" is the representation, the map, which uses the physical variables as its symbols. By searching and identifying the physical variables the students effectively develop a representation and the more effective the representation, the easier will be their learning process.

For instance, once the characteristic features of the ICM X-ray spectra are identified through a group activity (continuum and line emission, continuum emission appears to have an exponential cut-off at a given frequency), the naturally following question is, "What is the origin these observed features?" This question can be further refined into, "What are the physical properties of the emitting medium which determine the observed feature (e.g., the exponential cut-off)?" Through the other practices, as described below, the students can arrive at an effective representation based on an atomic scale model of free electrons which are slightly decelerated in their journey by the interaction with free protons (producing *Bremsstrahlung* radiation). In this model, the characteristics of the observed radiation are mainly determined by just two physical variables: volume density and velocity (which is represented in terms of "temperature", as discussed below in 3.2.3) of electrons and protons.

### 3.2.3 Making assumptions

Searching for the relevant physical variables in a complex system, through the construction of a "physical model" as seen above, often requires dealing with a large number of variables and physical processes. Dealing with such complexity is often so daunting that it can hamper the possibility to continue through the journey. The ability to make assumptions is a fundamental skill without which very few research activities would be possible.

Assumptions are thus central in research as much as in every learning activity. They can be categorized in at least two broad types: i) "underlying assumptions" which deal with our (known) ignorance of a part or some properties of the system and which allow us to continue our model construction in absence of the necessary knowledge, and ii) "simplifying assumptions" which make the construction of the physical model or mathematical calculations easier to solve and which could be verified *a posteriori*.

For instance, in the construction of the physical model which could explain the continuum emission from the ICM, the students realize that the velocity of the electrons is one of the relevant physical variables. But what are the velocities of the electrons? Do they all have the same velocity? Or they have a "distribution" of velocities? In analogy to other systems studied and observed in other "experiments" on Earth, the students could make the underlying assumption of a particular velocity distribution, i.e., the Maxwell-Boltzmann distribution, which characterizes the velocity as a function of another convenience variable defined as "temperature". We have however no way to directly verify if the electrons follow a Maxwell-Boltzmann distribution in the ICM *a priori*. Thanks to this assumption, however, the students can build a model — based on this assumption — which then can be applied to derive the density and "temperature" of the ICM. All the results are, however, dependent on this assumption.

In the construction of the model, the students are also invited to think about simplifying assumptions. One possibility is to consider the ICM as being composed of only free electrons and free protons (i.e., "hydrogen-only" composition and gas fully ionized). This allows a great simplification in the construction of the model and could be verified *a posteriori* from the data itself. Through a subsequent group activity, the students verify that, although not fully correct, the "cost" (in terms of "incorrectness" of the representation) versus "benefit" (simplification of the problem) of this assumption is what makes it acceptable.

The students are almost never directly exposed to this practice in their prior courses and in their exercise classes. Indeed, in their previous experiences the assumptions are either already given by the lecturer or in the exercise itself, or they are implicit in the problem. For this reason, this practice is central in the course design. In particular, to make the assumptions *visible* the following practice is used: at the beginning and during every investigation, the students are invited to divide their notes in two different areas, one of which is reserved for the list of assumptions they are making through their investigation. The students are constantly invited to think about the assumptions they are making in terms of their "plausibility", discussing them also with their colleagues. In particular, the students are invited to consider previous situations they have encountered, such as "equilibrium" situations in which physical processes balance each other in terms of rates or quantities. As for any other practice, trial and error strategies, although time consuming, are the most effective for learning "how to make assumptions". Students are thus invited to try continuing their model construction with a given assumption and then verify it at the first possible point in their activity. By making assumptions visible, the students also learn how much our "representations" or models are limited and that results obtained by frontier-research fields are never absolute, but instead are contingent upon a (sometimes very large) set of assumptions. By turning the problem upside down, they also realize that frontier-research can be seen as a *test* of our assumptions. When observations do not match our expectations, we must check and possibly change our assumptions, which, in a broader context represent also an ethical exercise (in additional to its logical and physics aspects).

### 3.2.4 Making testable hypotheses and predictions

Although the boundary between assumption and hypothesis in current language may be quite blurred due to the fact that they are often practiced together, in the design of the course these two terms have been used to represent distinct practices. In particular, in the course design, a hypothesis represents, e.g., a particular atomic or radiation process which is being tested through the comparison with the observational data. For instance, in the study of the ICM X-ray emission, a possible hypothesis is: "the continuum emission is due to the interaction between free electrons and free protons (*Bremsstrahlung*)". As such, while questions are the driver of the investigation, the hypotheses are the motor and milestones of the inquiry journey. Students' prior experiences with hypotheses are often indirect: they are usually given by the lecturer or by the exercises themselves. The ability to formulate hypotheses is a fundamental skill required for any authentic scientific research experience, and for this reason this practice is present in all the activities in the course. In particular, for every activity, hypotheses are *always* proposed and formulated by the students themselves in order to give them ownership of the learning process.

The facilitation strategy focuses on the concept of "testable hypotheses" which share similar characteristics of "refined questions" discussed above. After the students have observed and identified the main features of some dataset ("as they are"), they formulate hypotheses or "physical explanations" for the observed phenomena. During the activity, the students are invited to collect and *make visible* their hypotheses in their notebook, through words, sketches or in mathematical language. Making learning visible is also essential for Assessment (an ISEE Theme) by the facilitator and for self-assessment. In order to help the students in the choice of possible hypotheses to consider, one facilitation strategy is to ask them, "How would you test this hypothesis?" This helps the students focus again on the available material, including the observable variables and possible connections with a physical model described by physical variables.

The investigations are designed in any case to leave space for the "unexpected", i.e., for results which imply something that was not even "hypothesized". These are often the most important discoveries in science. For instance, after testing and verifying the hypothesis that the ICM X-ray emission is consistent with *Bremsstrahlung* radiation, the students can apply their model and find that the implied gas temperatures are extremely high, on the order of 10 million degrees Kelvin (!). By investigating why the gas is so hot (given a set of assumptions and a given physical model which the students make on their own, supported by facilitation) they invariably arrive at a similar possible conclusion: much more matter than what is visible must be present in order for the gas to be so hot! This unexpected result, which can be confirmed through several other independent experiments, could imply the presence of "dark matter", i.e., matter which interacts only gravitationally without producing radiation: a major discovery which emphasizes once again our *ignorance*, making space for more research, inquiry and future discoveries.

Together with the sense of discovery (often producing an "aha!" moment for the students which is also very rewarding for the facilitator), the activities are also designed to stimulate the sense of wonder connected to the study of the universe. For instance, a part of the activities is always reserved as a space for the students to *realize* some of the properties that they are finding by comparing them with their everyday or previous experiences. For instance, if a Galaxy Cluster would fit in a classroom, in this scale our planet would be smaller than a proton (!). Moreover, the densities of matter outside of galaxies, such as the ICM, which the students find and study through their emission, are 6 orders of magnitude below the best vacuum we can make in our laboratories on Earth (!). It is much emptier than "empty space" but the size of these structures implies a mass which is much larger than thousands of billions of Suns.

### 3.2.5 Reducing complex problems to smaller units

From the formulation of a refined question or hypothesis to the development of a "physical model", the students always find themselves in complex situations reflecting the complex nature of reality. In many cases, however, complex problems can be split and studied in smaller units, which are then joined together at a later stage without loss of accuracy. Activities involving radiation processes are ideal laboratories for this practice. Indeed, thanks to the additive nature of radiation, several emission processes could be split to the smallest possible units: individual protons and electrons interacting with each other. The majority of the activities are therefore designed as a series of possible steps in which students have the possibility to add elements one at a time.

As an example, let us consider again the activity on the origin of ICM X-ray emission. During this activity, through facilitation, the student can first develop an understanding of the emission produced by one individual electron interacting with one proton (as a function of the relevant physical variables: distance between particles and electron velocity). Then they can consider multiple electrons interacting with multiple protons (a new physical variable emerges: density). Multiple electron velocities, through the assumption of a velocity distribution, can be considered (another new variable emerges: temperature). Added together, these elements can explain the observed properties of the X-ray continuum emission. By traveling this (facilitated) journey on their own, students take once again ownership of this important practice.

Moreover, during each step, learning opportunities of foundational scientific content arise, which would not have been possible without splitting the problem into smaller parts. For instance, by considering the interaction between one electron and one proton, the student can realize fundamental aspects of the nature of light and spectra: a sudden change in the electric field is able to produce emission on a large spectrum of frequency (this step is facilitated through a discussion on *Fourier Analysis*, to which students have been exposed in previous courses, in a graphical way). By considering a distribution of velocities, the student can realize the *nature* of one of the variables which are used in their everyday life, i.e., "temperature". These elements are practiced again and again in the other activities, which are all constructed in a similar way, although their content framework is different.

### 3.2.6 Sharing

The communication of the results of the inquiry, at any stage, is one of the fundamental steps in the learning process. In order to communicate ideas or concepts to others, a systematic organization of the inner dialogue or "thought" is needed. This practice allows the consolidation of ideas and is therefore useful as much for the speaker as it could be for the listener (if properly organized). Students are not used to this practice in their previous frontal-lecture experiences as listeners. For this reason, the "plenary sharing" phase, in which each group reports to the rest of the classroom, is always included in all the activities of the course and it is facilitated through prompts, if needed. This phase typically corresponds to the culminating stage of the inquiry activity which has been defined by an overall driving question and it is also the moment in which different groups can compare their results, finding similarities or differences. Because the facilitator is not the owner of knowledge, the only way in which students can assess their results, e.g., in terms of quantitative measurements, is through comparison with the other groups, as would be the case for an authentic research experience. Disagreement between numerical results is seen therefore not necessarily as a "negative" thing since it allows the students to learn how to find possible errors on their own. Because assessment and consolidation of foundational content and practice happens through the facilitation of the group activity, disagreement on these aspects is not present at this stage.

Students are invited to prepare their "sharing" following the same structure of their investigation: listing assumptions, hypotheses, providing a sketch or representation (in any form) of their physical model and listing the individual steps of their investigation. The presentation within an individual group is organized by the students themselves who are, however, encouraged by the facilitator to equally share and to all be present "on stage" during their presentation. This allows the possibility of contributing independent of a student's level of confidence in "speaking in front of an audience". It also allows students to acquire relevant skills according to each student's individual pace. "Sharing" is made by students for the other students. As such, the facilitator's role here is to chair the presentation without intervening and to help collect questions from the audience, giving complete ownership of the process to the students.

The results of this "sharing phase" are then recollected at the beginning of the following class in a recurring slide (with title: "What you have learned last class") in which the facilitator reports the results of the students (who are indeed the owners of the results). The material is organized in order to provide a clear future reference for the students, to be compared with their own notes and material.

## 4. Discussion

Once put into practice, has the course design achieved the expected goals? In this section, I briefly discuss the course's outcomes and limitations as seen by different points of view, starting with my direct experience as "designer" and facilitator.

One striking feature of a course designed as described in this article is that every year it is different because the (yearly changing) students themselves are the active agents of the course. This means that every aspect of the course, as the learning journey unfolds, must adapt to the students rather than the opposite (which is the norm for monologic lecture-based courses that can be invariably the same every year). This is at the same time challenging but extremely interesting and thus rewarding: every class is a new experience and a new learning opportunity for the facilitator. Like for any authentic learning experience, facilitation requires practice. Although there is of course literature and previous experiences collected by other facilitators on this topic, it cannot be really fully learned by reading a book or a manual: the path must be walked to be learned. Thus, although the students are not realizing it, through their experience together, the facilitator learns at least as much as they do — sometimes struggling as much as they do when they feel "stuck" in the middle of an activity, and, most importantly, enjoying as much as they do their "aha!" moments.

In the design and preparation of the activities, given their nature, it essential to try to *practice* them as students would do, trying to anticipate possible problems or issues during the journey and allowing multiple paths. Because of this, preparing such a course is much more time-consuming than a monologic lecture class. In my case, the time spent in preparation has been almost always rewarded in the classroom, but this might not fit into the expectations of everyone.

There are of course intrinsic limitations (i.e., not due to a poor design) of this teaching approach, like for any method, with respect to other methods. In particular, given its dialogic nature, this method necessarily requires a high facilitator to student ratio. Current large universities, especially at the BSc level, are simply not designed in this way: what kind of dialogue would be possible in a lecture room with one lecturer and 300 students? The small number of students in an MSc course has without doubt favored a possible dialogic approach. In particular, at ETH, I have been supported in an excellent way by a more junior co-facilitator and only rarely one facilitator had the responsibility for more than two groups of students (which made facilitation time-efficient). At UniMiB, where I have done the facilitation on my own, I regularly had to facilitate 4 groups. This has sometimes proved to be a challenge, making the advancement of groups through the most difficult activities (requiring more intense facilitation) quite uneven. In order to cope with this situation, in addition to further simplifying the activities from unnecessary complications, I have re-designed some activities providing more

"scaffolding", which helped students advance without the "constant" presence of the facilitator.

An apparent limitation (as seen through the lenses of a monologic lecture-based course) is the fact that much less "content" can be covered using a dialogic, inquiry-based approach. I would argue that this is instead an advantage: it allows the design of the course to focus on what is really important (including "practices") instead of purely "transferring" information (which is then not retained) — something not particularly useful in our current society where information can be easily accessed everywhere.

As is the case for any dialogue, linguistic barriers can be sometimes a limitation, especially in situations including students with different backgrounds: not all of them may be confident enough to speak and express their ideas in a foreign language, i.e., in English (a foreign language for almost all the students attending the course). Also in this case, however, a limitation can become an advantage, if such an experience is seen as a way to train the students in expressing themselves in a language that they would need in any case in order to become active members of an international scientific community.

Very low to null course drop rates are the first observed outcomes on the students' side. Once the initial skepticism about a completely new (for them) course format is overcome, the students are able to realize through the course activities that their active participation is essential for their success in the course. From their feedback, the students particularly appreciate that this format gives them the opportunity to talk and discuss with each other, to become a real group working together rather than isolated individuals in a classroom. The feeling of being part of a group reinforces their intrinsic motivation to come to the classroom and to participate. An important strategy to achieve this is to form random and new groups for each activity in such a way that everyone has worked with everyone else at least a few times. Facilitation and the sharing phase are also essential to stimulate participation of every group member so that everyone feels integrated. This is particularly challenging in classrooms (e.g., in my experience at ETH) including students with very different backgrounds, and close attention must be paid to group dynamics and plenary sharing in these situations.

Making learning visible through dialogic interaction greatly facilitates assessment during each activity. Learning outcomes are thus constantly assessed both by the facilitator and by the students themselves. By realizing that they are effectively learning something, students' positive feeling towards active participation can be reinforced. I have heard a few times in these years the students exclaiming "I finally understand physics!", which is one of the most rewarding moments also for the facilitator. However, particular care is needed in the activity design, in order to avoid goals that are too difficult to achieve and an "unsuccess" (which is also part of real research activities). This would produce instead frustration and negative feelings in the students. At the same time, an activity which "appears" too easy would also not be particularly rewarding for the students. On the facilitator side, this balance can be often learned only by experience.

Despite all the efforts on both the facilitator and student sides, there are of course situations in which a part of the content or practice goals are not finally achieved by the end of the course by all students. This is also the reason why the final assessment (which uses a detailed rubric with the main content and practice goals describe above) is structured as a way to provide point-by-point feedback rather than as a "standard" exam. Following the long-standing and often forgotten tradition highlighted in the Introduction, the final goal of the course is indeed to provide a way for the students (and to the facilitator) to improve themselves, as researchers of the nature of things around us, and as members of a community, as well as individuals.

# 5. Summary and concluding remarks

Can learning and teaching be a *truly formative* or *transformative* experience that would help us see the world around us (and ourselves) in a new way? I have of course no answer to this question since answers are not the goal of this article. As the ancients would say, the path must be walked to be learned.

In the first part of this article (Section 1), with the help of contemporary historians of ancient philosophies (including in particular Pierre Hadot), I have tried to show how such a path has been travelled by several of the most important teachers of Antiquity for which we have written sources, both in Europe and in Asia. In particular, these teachers emphasized through their teaching the pivotal importance of "practice" and dialogic learning for a truly formative experience. After many centuries, the importance of these themes has been (once again) rediscovered in the last decades and applied to modern contexts, e.g., through inquiry learning approaches such as the ISEE Inquiry Framework, in which foundational scientific content and practice are interwoven.

I have described here how, through this framework, a series of MSc courses on the topic of "Cosmic Structure Formation" have been designed and implemented in a European university environment (Sections 2 and 3). The courses are structured as a series of facilitated and dialogic experiences which mirror authentic scientific research, including all aspects which are usually not *practiced* in lecture-based courses. These include the following: generating and refining questions; making and testing assumptions; developing one's own research path by making and testing hypotheses; and sharing, explaining and justifying ideas and results with peers.

These practices are interwoven with foundational scientific content, which has the potential of transforming the students' view of the world around them going from the "extremely small" of the atomic and subatomic world to the "extremely large" of the universe — worlds which are deeply interconnected to each other. One of the foundational content goals of the courses has been to let the students (re-)discover these connections by using radiation processes on atomic scales as a mean to unravel the physical properties of the largest structures in the universe.

In the last part of the article (Section 4), I discussed some of the most rewarding and challenging aspects of the courses' implementation both on the facilitator's and students' sides. By traveling the path together, as a part of a community of learners which included the facilitator, we have all learned very much from each other.

# Acknowledgements

I am grateful to Gabriele Pezzulli who has shared with me the experience of designing and facilitating the first courses on "Cosmic Structure Formation" together at ETH and without whom these courses would not have been possible.

The PDP was a national program led by the UC Santa Cruz Institute for Scientist & Engineer Educators. The PDP was originally developed by the Center for Adaptive Optics with funding from the National Science Foundation (NSF) (PI: J. Nelson: AST#9876783), and was further developed with funding from the NSF (PI: L. Hunter: AST#0836053, DUE#0816754, DUE#1226140, AST#1347767, AST#1643390, AST#1743117) and University of California, Santa Cruz through funding to ISEE.